\documentclass[twocolumn,preprintnumbers,notitlepage,superscriptaddress,amsmath,10pt,aps,pra]{revtex4-1}
\usepackage{graphicx}
\usepackage{color}
\usepackage{hyperref}
\usepackage{subfigure}
\usepackage{graphicx}
\usepackage{type1cm}
\usepackage{eso-pic}
\usepackage{color}
\usepackage{amsmath}
\usepackage{amssymb}
\usepackage{physics}
\usepackage{url}
\usepackage[sort&compress]{natbib}

\begin{document}

\title{Incomplete quantum oblivious transfer with perfect one-sided security} %


\author{David Reichmuth}
\address{IPaQS, Heriot-Watt University, Edinburgh, UK}
\author{Ittoop V. Puthoor}
\address{School of Computing, Newcastle University, Newcastle upon Tyne, UK}
\author{Petros Wallden}
\address{School of Informatics, University of Edinburgh, Edinburgh, UK}
\author{Erika Andersson}
\address{IPaQS, Heriot-Watt University, Edinburgh, UK}

\begin{abstract}
Oblivious transfer is a fundamental cryptographic primitive which is useful for secure multiparty computation. There are several variants of oblivious transfer. We consider 1-out-of-2 oblivious transfer,  where a sender sends two bits of information to a receiver. The receiver only receives one of the two bits, while the sender does not know which bit the receiver has received. Perfect quantum oblivious transfer with information-theoretic security is known to be impossible. We aim to find the lowest possible cheating probabilities. Bounds on cheating probabilities have been investigated for ``complete" protocols, where if both parties follow the protocol, the bit value obtained by the receiver matches the sender's bit value. We instead investigate incomplete protocols, where the receiver obtains an incorrect bit value with probability $p_f$. We present optimal non-interactive protocols where Alice's bit values are encoded in four symmetric pure quantum states, and where she cannot cheat better than with a random guess. We find the protocols such that for a given $p_f$, Bob's cheating probability $p_r$ is as low as possible, and vice versa. Furthermore, we show that non-interactive quantum protocols can outperform non-interactive classical  protocols, and give a lower bound on Bob's cheating probability in interactive quantum protocols. Importantly for optical implementations, our protocols do not require entanglement nor quantum memory. 
\end{abstract}

\maketitle

\section{Introduction}

1-out-of-2 oblivious transfer (OT) is a cryptographic primitive where a sender Alice holds two bits, and a receiver Bob  receives one of them. The receiver should not know the other bit, and the sender should not know which bit the receiver obtained. A dishonest sender Alice will attempt to find out which bit Bob obtained, and a dishonest receiver Bob will attempt to learn both bit values. Apart from 1-out-of-2 oblivious transfer \cite{Wie83, evengoldlemp}, there are other variants. In Rabin oblivious transfer \cite{Rabin}, for example, a single bit is either received or not. Oblivious transfer is important not the least because it is universal for multiparty computation~\cite{OTuniversal}.

Unlike for quantum key distribution, information-theoretically secure perfect quantum oblivious transfer is impossible \cite{Mayers,LoNogo}. It does become possible with restrictions on quantum memory for cheating parties~\cite{DFSS05}.
Variants of oblivious transfer where the parties are constrained by special relativity are also possible~\cite{Garcia}. 
While perfect quantum oblivious transfer is impossible without restrictions on cheating parties, 
quantum protocols with bounded cheating probabilities for the (otherwise unrestricted) parties are still possible.
Currently known protocols however do not achieve cheating probabilities that are tight with existing lower bounds. The known bounds also hold specifically for ``complete" protocols \cite{Sikora11, Chailloux13, Chailloux16, Ryan}, where completeness means that the protocol always works correctly if the parties are honest. For such protocols, a lower bound on the greater of Alice's and Bob's cheating probabilities is 2/3 in general \cite{Chailloux16, Ryan} and $\approx 0.749$ if symmetric pure states are used~\cite{Ryan}. Another variant  of oblivious transfer is XOR oblivious transfer (XOT) where the sender has two bits and a receiver obtains either the first bit, the second bit, or their XOR. Using pure symmetric states, the XOT protocol has been demonstrated to be an optimal protocol having lower cheating probabilities than the classical XOT protocols~\cite{Stroh}.

We will here instead examine non-interactive incomplete protocols for 1-out-of-2 quantum oblivious transfer. Non-interactive means that there is no back-and-forward communication, either classical or quantum; the protocols have a single step, where one party sends classical and/or quantum information to the other party, who measures what is received. Some of our results also provide bounds valid for interactive protocols.
``Incomplete" means that a protocol might fail even when both parties follow the protocol. Here protocol failure will mean that Bob obtains an incorrect bit value. There are several reasons to investigate protocols that can fail. It turns out that a non-zero failure probability means that cheating probabilities can be lowered. There is also a connection to quantum random access codes~\cite{Wie83, QRAC}, as will be discussed at the end; incomplete oblivious transfer can be seen as a generalisation that ``interpolates" between random access codes and complete oblivious transfer. Noise and imperfections will also often lead to a non-zero failure probability in implementations.

Specifically, we investigate incomplete protocols for quantum oblivious transfer, where the sender Alice can never cheat any better than with a random guess. That is, the security against a cheating Alice is ``perfect". For a given protocol failure probability $p_f$, the goal is then to minimize Bob's cheating probability $p_r$, and vice versa. Bob's cheating probability is the probability that he correctly guesses both of Alice's bit values.

We discuss the definition of protocol failure in Section \ref{sec:fail}, non-interactive classical protocols in Section \ref{sec:class}, and some general properties of incomplete quantum oblivious transfer in Section \ref{sec:gen}. In Section \ref{sec:sym}, we find the lowest possible cheating probability for receiver Bob for a given protocol failure probability for protocols using symmetric pure states, and in \ref{sec:opt}, we give  examples of optimal quantum protocols of this type. The example protocols are feasibleto realize with current technology, for example using photons. We finish with a discussion.

\section{Definition of protocol failure}
\label{sec:fail}
``Failure" could in principle mean either that Bob obtains an incorrect bit value, or that he fails to obtain any bit value at all. An incorrect bit value could be due to noise or randomness that is in principle avoidable (but might be intentional), or, in a quantum protocol, to the fact that non-orthogonal quantum states cannot be perfectly distinguished from each other. Bob could fail to obtain any bit value at all again because of randomness or noise, e.g. if no detector on Bob's side registers anything. Or, in a quantum protocol, if Bob is for example using an unambiguous quantum measurement to determine one of Alice's bit values, then such a measurement might give an inconclusive result, meaning that Bob fails to obtain a bit value even if the implementation is perfect.

We will concentrate on incomplete protocols where Bob always obtains a bit value, which is incorrect with probability $p_f$, and where Bob generally does not know if his bit value is correct or not. The other type of incomplete protocol would be if Bob sometimes fails to obtain any bit value at all (in addition, when he does obtain a bit value, it might be guaranteed to be correct, or might be correct with a probability less than one). If Bob fails to obtain a bit value at all, then this will be obvious to Bob. If both parties are honest, they could agree to simply repeat the protocol until a bit value is obtained by Bob (with Alice using new bit values each time). Taken together, if the bit value that Bob in the end obtains is guaranteed to be correct, this would then simply be another instance of a complete protocol for quantum oblivious transfer. If the bit value in the end obtained by Bob's is sometimes incorrect, then the protocol is of the type we consider.
Apart from repeating the protocol until success, the parties could also agree that if Bob fails to obtain a bit value, he makes a random guess to produce a bit value. In this case Bob again obtains an incorrect bit value with some probability. This is therefore again a protocol of the type we will be considering. Generally, Bob's lowest failure probability $p_f$ will be achieved when he is never certain that the bit value he received is correct (because Bob's optimal measurement will be a minimum-error measurement).

To summarize, we will consider the case where Bob always obtains a bit value, which is incorrect with probability $p_f$, and where Bob does not know when the protocol has failed (or at least does not always know). In the quantum protocols we will examine, Alice does not know whether the protocol has succeeded or failed either.

Cheating probabilities could be calculated either as overall cheating probabilities, whether the protocol has failed or not, or as conditional cheating probabilities, conditioned on the protocol failing or succeeding. In the quantum protocols we consider, the honest parties do not know whether a particular run of the protocol has failed or not. It  therefore seems most natural to consider overall cheating probabilities. In the classical non-interactive protocols below, on the other hand, Alice may know whether or not a bit value is correct (but not whether Bob has received it). This can be seen as an advantage for Alice. We will nevertheless compare overall cheating probabilities for classical and quantum protocols.

\section{Classical non-interactive protocols}
\label{sec:class}

As always in quantum cryptography, we will need to compare our quantum protocols to classical protocols, and hence we will need to know how low cheating probabilities can be in classical protocols.
First, recall that in any complete protocol for 1-out-of-2 oblivious transfer (quantum or classical), either party can always successfully cheat with a probability of $1/2$ using a random guess. 
Furthermore, in protocols with information-theoretic (as opposed to computational) security, if one party can successfully cheat only with probability $1/2$, then the other party can necessarily cheat with probability 1~\cite{Chailloux16}. This holds both for ``classical" and for quantum protocols which are complete. A classical protocol achieving this is where Alice sends both bits to Bob, who (if he is honest) reads out only one of them. Alice's cheating probability is then 1/2 (her probability to correctly guess which bit Bob obtains) and Bob's cheating probability is 1 (his probability to obtain both of Alice's bits). Conversely, if Alice sends one of the bits to Bob and ``forgets" which one she sent if she is honest, then Alice can cheat with probability 1 and Bob can cheat with probability 1/2.

Let us now consider incomplete classical protocols which fail with some probability $p_f$, where the sender can only cheat with probability $1/2$, to see how much the receiver's cheating probability can be lowered. Let us start with an example. Suppose that Alice sends two bits to Bob, but that she (or a ``noisy environment" that Bob cannot control or access) flips one of her bit values with some probability, and leaves the other one unchanged. Each bit is equally likely to be the flipped one. Suppose that the first bit is flipped and the second bit left correct with probability $p\le 1/3$, and the first bit left correct and the second one flipped also with probability $p$. Both bit values are left correct with probability $1-2p\ge 1/3$ (and it never happens that both bit values are flipped). The failure probability is then $p_f = p$, which is the probability that an honest Bob is unlucky and reads out a flipped bit value. Bob's cheating probability if he reads out both bits is $1-2p = 1-2p_f $.

For a non-interactive classical protocol, this turns out to be optimal, in the sense that for a given $p_f\le 1/3$, Bob's cheating probability $p_r$ is as small as it can be, keeping Alice's cheating probability equal to $p_s=1/2$. 
Let us consider a general classical non-interactive protocol where Alice sends some information to Bob, who then reads out one bit if he is honest, and both bits if he is dishonest. ``Reading out" may involve some information processing; we are placing no computational restrictions on either Alice or Bob. 
Classical information can be copied, and therefore, for example, the probability that if Bob chooses to read out the first bit value, it is correct, is independent of whether or not he also chooses to read out the second bit value. (This would not be the case in a quantum protocol, where Bob generally, if he learns about a particular property of a quantum state, will be able to obtain less information about something else.)

Let us denote the probability that both bit values can be correctly read out by Bob by $c$, the probability that the first bit value is correct and the second is wrong by $p$, the probability that the first bit value is correct and the second is wrong by $q$, and the probability that both bit values are wrong by $w$. It holds that $c+p+q+w=1$. 
We can without loss of generality assume that $c$ is larger than $p, q$ and $w$ (if this is not the case to start with, then Bob can flip bit values to make it so). Bob's cheating probability will then be $p_r=c$. An honest Bob's success probability to obtain one bit value correctly, if he randomly reads out either the first or the second bit, is $1-p_f=c+(p+q)/2$. The probability that his bit value is wrong is $p_f=w+(p+q)/2$. It therefore holds that  
\begin{equation}
p_r=c= 1-w-q-p = 1-2p_f+w.
\end{equation}

In order to minimize Bob's probability to cheat, for a given $p_f$ for an honest Bob, Alice should clearly choose $w$ as small as possible. This means that in addition to $c={\rm Max}(c,p,q,w)$, Alice will ensure that $w={\rm Min}(c,p,q,w)$. Other choices are possible, but will lead to a higher cheating probability for Bob, for a given protocol success probability, and we (and Alice) are interested in the {\em lowest} possible cheating probability for Bob. If $c \ge 1/3$, then Alice can choose $p, q, w$ so that $w=0$, and it then holds that $p_r= 1-2p_f$. Generally,  when $0\le p_f\le 1/3$, for classical non-interactive protocols it therefore holds that
\begin{equation}
\label{eq:classbound1}
p_r\ge 1-2p_f.
\end{equation}
If $c\le 1/3$, then the smallest $w$ Alice can choose is $w=1-3c$, by picking $p=q=c$. In this case, the protocol failure probability $p_f=1-2c=1-2p_r$. When $p_f \ge 1/3$, it therefore holds that Bob's cheating probability is bounded as
\begin{equation}
\label{eq:classbound2}
p_r\ge \frac{1}{2}(1-p_f).
\end{equation}
A receiver who performs the protocol as if they are honest, but then randomly guesses the bit value they do not obtain, will correctly guess both bit values with probability $p_r=\frac{1}{2}(1-p_f)$. Equation \eqref{eq:classbound2} therefore means that for $p_f \ge 1/3$, this is Bob's best cheating strategy.

This combined bound in equations \eqref{eq:classbound1} and \eqref{eq:classbound2}, for non-interactive classical protocols where the sender can only cheat with probability $p_s=1/2$, will serve as a benchmark for our quantum protocols. It turns out that for corresponding non-interactive quantum protocols, the cheating probability for the receiver Bob can be {\em less} than in classical protocols.

\section{Incomplete quantum protocols}
\label{sec:gen}

We will now consider quantum oblivious transfer with a failure probability $p_f$, where the sender can only cheat with probability $p_s=1/2$, that is, no better than using a random guess. As mentioned above, we define the sender's and receiver's cheating probabilities as their overall probability to cheat successfully, independently of whether the protocol fails or not. This makes sense since the parties will not know whether or not the protocol has failed. (In complete quantum protocols, dishonest parties will also not always know whether the information they have obtained is correct, if they maximize their cheating probabilities.)

We will consider non-interactive protocols where a quantum state is sent only once from sender to receiver, who makes a final measurement.
In general, protocols for quantum oblivious transfer can have several steps where quantum states are sent back and forth, such as in the protocol by Chailloux et al.~\cite{Chailloux13}, and as in the general framework in \cite{Ryan}. Some of our results generalize to such protocols. In particular, a dishonest party can always cheat by honestly performing the protocol until the last step, and then cheating by altering their final measurement. The cheating probability for this particular strategy then gives a lower bound for their general cheating probability. That is, lower bounds for cheating in non-interactive protocols also give lower bounds for cheating in 
interactive protocols.

In 1-out-of-2 oblivious transfer Alice has two classical input bits $x_0, x_1$, and Bob has one input bit $c$. Usually one considers the case where Alice and Bob select their bit values randomly.
Suppose that an honest sender Alice encodes her two classical bit values in one of the quantum states $\sigma_{00}, \sigma_{01}, \sigma_{10}, \sigma_{11}$, where the subscripts indicate Alice's bit values. 
She sends this quantum state to Bob, who (if he is honest) selects between making a measurement to learn either the value of the first bit ($c=0$), or a measurement to learn the value of the second bit ($c=1$), with equal probability. 
Bob choosing between two measurements, and the fact that there is no further communication with Alice, will guarantee that Alice's overall cheating probability is $1/2$. By no-signalling, she cannot tell which measurement Bob is choosing, or indeed whether Bob has made a measurement at all~\cite{rimini}. She therefore cannot cheat better than with a random guess. Bob's cheating probability on the other hand can be higher than $\frac{1}{2}(1-p_f)$, which he would achieve with a random guess. Conversely, if Alice should be able to cheat no better than with a random guess, then she should not be able to distinguish between Bob obtaining the first or the second bit value. It follows that whatever Bob does can, from Alice's point of view, be described as him choosing between two different generalized measurements, depending on which bit value he wishes to obtain. 

We will now examine how well the protocol works if both parties are honest.
If Bob is honest and wishes to learn the value of the first bit, then he performs a measurement to distinguish between the sets of states $S_{00}=\{\sigma_{00}, \sigma_{01}\}$ and $S_{01}=\{\sigma_{10}, \sigma_{11}\}$, where all states are equiprobable. This is the same as making a measurement to distinguish between the equiprobable states $\frac{1}{2}\sigma_{00}+\frac{1}{2}\sigma_{01}$ and $\frac{1}{2}\sigma_{10}+\frac{1}{2}\sigma_{11}$. The minimum-error measurement that distinguishes between two states $\rho_0$ and $\rho_1$, occurring with probabilities $p_0$ and $p_1$, has the error probability
\begin{equation}
p_{\rm err} = \frac{1}{2}(1-{\rm Tr}|p_0\rho_0-p_1\rho_1|).
\end{equation}
The measurement is a projection in the eigenbasis of $p_0\rho_0-p_1\rho_1$. 
The probability that Bob obtains an incorrect value for the first bit, if he uses a minimum-error measurement, is therefore
\begin{equation}
p_{f,0} = \frac{1}{2}(1-\frac{1}{4}{\rm Tr}|\sigma_{00}+\sigma_{01}-\sigma_{10}-\sigma_{11}|).
\label{eq:pf0}
\end{equation}
If Bob wishes to learn the second bit, then he performs a measurement to distinguish between the sets of states $S_{10}=\{\sigma_{00}, \sigma_{10}\}$ and $S_{11}=\{\sigma_{01}, \sigma_{11}\}$, where again all states are equiprobable. This is equivalent to distinguishing between the equiprobable states $\frac{1}{2}\sigma_{00}+\frac{1}{2}\sigma_{10}$ and $\frac{1}{2}\sigma_{01}+\frac{1}{2}\sigma_{11}$. Bob's minimum-error measurement gives an incorrect value for the second bit with the probability
\begin{equation}
p_{f,1} = \frac{1}{2}(1-\frac{1}{4}{\rm Tr}|\sigma_{00}+\sigma_{10}-\sigma_{01}-\sigma_{11}|).
\label{eq:pf1}
\end{equation}
On average, Bob's probability to obtain an incorrect bit value is
\begin{equation}
p_f = \frac{1}{2}(p_{f,0}+p_{f,1}).
\end{equation}
If Bob's failure probability is independent of whether he tries to obtain the first or the second bit, then $p_{f,0}=p_{f,1}=p_f$.

We already remarked that Alice's best cheating strategy is a random guess. If Bob wishes to cheat, then he needs to distinguish between all four of Alice's states. His optimal measurement for this necessarily gives the wrong result at least as often as either of the above two measurements where he learns only one bit value. This means that Bob's cheating probability $p_r$, defined as the probability that he can correctly guess both of Alice's bit values, is generally strictly smaller than either of $p_{f,0}, p_{f,1}$ or $p_f$ above. That is, it holds that
\begin{equation}
p_r \le 1-p_{f,0},~ p_r \le 1-p_{f,1} {\rm ~~ and~~}p_r \le 1-p_f,
\end{equation}
and the inequalities are strict except in special cases (which would correspond to poorly designed protocols). 
However, for a quantum OT protocol to have a lower cheating probability for Bob than the best non-interactive classical protocol, we would want $p_r$ to be lower than the bounds in Eqns. \eqref{eq:classbound1} and \eqref{eq:classbound2}.

We can also make further statements about $p_f$ and $p_r$. 
From equations \eqref{eq:pf0} and \eqref{eq:pf1}, if Bob is selecting between learning either the first or the second bit value, 
then his failure probability $p_f$ will be nonzero unless the states $\sigma_{00}, \sigma_{01}, \sigma_{10}, \sigma_{11}$ all have distinct supports, and can therefore be perfectly distinguished from each other. That is, as was already known, $p_f=0$ necessarily implies that  Bob's cheating probability $p_r=1$, if we also demand that Alice should not be able to cheat any better than with a random guess.

Moreover, that $p_f=0 \wedge p_s=1/2$ implies $p_r=1$ holds true whether or not $c$ is actively chosen by Bob. In so-called semi-random protocols, Bob does not select between two courses of action during the protocol depending on his input $c$, but instead obtains his value of $c$ randomly as an output from the protocol~\cite{Ryan}. Without loss of generality, whether or not Bob actively chooses $c$, we can assume that Bob obtains both $c$ and the bit value $x_c$ using a measurement that he defers to the end of the protocol. Bob's final measurement then has four outcomes. We denote Bob's generalized measurement operators by $\pi^{i*}$ for $c=0$, and $\pi^{*i}$ for $c=1$, where $i=0, 1$ gives the value of $x_c$. For example, the measurement operator $\pi^{1*}$ corresponds to $c=0$ and $x_0=1$. Alice can cheat by using entanglement, so that she and Bob in the last step of the protocol share an entangled state. The requirement that Alice is unable to learn anything about $c$ means that the state she holds after Bob's measurement cannot depend on Bob's output $c$. From no-signalling~\cite{rimini}, this implies that $\pi^{0*}+\pi^{1*}= q \hat{\bf 1}$ and $\pi^{*0}+\pi^{*1}= (1-q) \hat{\bf 1}$, where $0\le q\le 1$. Bob's measurement is therefore equivalent to him making a random choice between learning the first bit with probability $q$ and the second bit with probability $1-q$. Equations  \eqref{eq:pf0} and \eqref{eq:pf1} then again mean that if $p_f=0 \wedge p_s=1/2$, then $p_r=1$ holds, also when Bob does not choose $c$ as an input, but obtains it as an output from the protocol.

When $p_f>0$, we can reason as follows. Let us assume that the same measurements Bob would use to learn either the first or the second bit is used instead when it is in addition known that Alice is sending either $\sigma_{00}$ or $\sigma_{11}$, with equal probability. In this case, learning either of the bit values determines the state. If $p_{f,0}$ does not depend on whether $\sigma_{00}$ or $\sigma_{01}$ was prepared, or whether $\sigma_{10}$ or $\sigma_{11}$ was prepared, it follows that
\begin{equation}
p_{f,0} \ge \frac{1}{2}(1-\frac{1}{2}{\rm Tr}|\sigma_{00}-\sigma_{11}|),
\end{equation}
where the RHS is the minimum error probability for the measurement that optimally distinguishes between the equiprobable states $\sigma_{00}$ and $\sigma_{11}$.
Similarly, if $p_{f,1}$ does not depend on whether $\sigma_{00}$ or $\sigma_{10}$ was prepared, or whether $\sigma_{01}$ or $\sigma_{11}$ was prepared, we have
\begin{equation}
p_{f,1} \ge \frac{1}{2}(1-\frac{1}{2}{\rm Tr}|\sigma_{00}-\sigma_{11}|).
\end{equation}
If both of these equations hold, then we obtain
\begin{equation}
p_f\ge \frac{1}{2}(1-\frac{1}{2}{\rm Tr}|\sigma_{00}-\sigma_{11}|).
\end{equation}
If $\sigma_{00}=\ket{\psi_{00}}\bra{\psi_{00}}$ and $\sigma_{11}=\ket{\psi_{11}}\bra{\psi_{11}}$ are pure, then we obtain
\begin{equation}
p_f \ge \frac{1}{2}(1-\sqrt{1-|\braket{\psi_{00}}{\psi_{11}}|^2}).
\label{eq:pfgbound}
\end{equation}
A similar argument can be used to derive
\begin{equation}
p_f\ge \frac{1}{2}(1-\frac{1}{2}{\rm Tr}|\sigma_{01}-\sigma_{10}|).
\end{equation}

Bob's optimal cheating strategy will be a minimum-error measurement for distinguishing between all four states $\sigma_{00}, \sigma_{10}, \sigma_{01}$ and $ \sigma_{11}$. In a general interactive protocol with more steps, the success probability of this measurement gives a lower bound for Bob's cheating probability.
We will use a bound by Audanaert and Mosonyi~\cite{AM14} for the minimum-error probability for distinguishing between a set of states $\{\rho_i\}$ with prior probabilities $p_i$,
\begin{equation}
p_{\rm err} \le \frac{1}{2}\sum_{i\ne j} \sqrt{p_ip_j}F(\rho_i,\rho_j),
\end{equation}
where the fidelity of two quantum states (sometimes called square root fidelity) is defined as
\begin{equation}
F={\rm Tr}[(\sqrt{\rho_0}\rho_1\sqrt{\rho_0})]^{1/2}.
\end{equation}
For two pure states this reduces to the absolute value of their overlap.
We denote the largest of the ``nearest-neighbour" fidelities between $\sigma_{00}$ and $\sigma_{01}$, between $\sigma_{00}$ and $\sigma_{10}$,   between $\sigma_{11}$ and $\sigma_{01}$ and  between $\sigma_{11}$ and $\sigma_{10}$ by $F$, and the larger of the
fidelities between $\sigma_{00}$ and $\sigma_{11}$, and between $\sigma_{01}$ and $\sigma_{10}$, by $G$. Bob's cheating probability then obeys
\begin{equation}
p_r \ge 1 -F-\frac{1}{2}G.
\label{eq:pbfgbound}
\end{equation}
If $p_f=0$, then $G=0$ is necessary, and the equation above reduces to the corresponding bound for complete protocols~\cite{Ryan}.

Equation \eqref{eq:pfgbound} (or the corresponding equation for $\ket{\psi_{01}}$ and $\ket{\psi_{10}}$, depending on which fidelity is larger) can be rewritten as
\begin{equation}
\sqrt{p_f(1-p_f)}\ge \frac{1}{2}G,
\end{equation}
which holds if the $\sigma_{ij}$ are pure states.
It then holds that
\begin{equation}
\label{eq:bobpurebound}
p_r\ge 1-F-\sqrt{p_f(1-p_f)}.
\end{equation}
From the above equations, it seems that a non-zero failure probability $p_f$ may allow for a lower cheating probability $p_r$ for Bob (lower than if $p_f=0$). The minimum of the RHS in \eqref{eq:bobpurebound} is obtained when $p_f=1/2$, giving $p_r\ge 1/2-F$, but $p_f=1/2$ corresponds to a random (and thus useless) result for an honest Bob. The expression is of course only a lower bound on Bob's cheating probability, but the lower bound is lower for non-zero $p_f$ and $G$. 

Alice's cheating probability remains equal to 1/2 independent of $F$ and $G$, since an honest Bob is randomly selecting between two measurements to learn either the first or the second bit value. For complete protocols, it generally holds that $p_s\ge (1+F)/2$~\cite{Ryan}, which together with $p_r\ge 1-F$ means that for complete protocols, Alice's and Bob's cheating probabilities obey the tradeoff relation $2p_s+p_r\ge 2$. For the incomplete (imperfect) protocols we are examining, with $p_s=1/2$, it instead holds that
\begin{equation}
2p_s + p_r \ge 2 - F- \frac{1}{2}G,
\end{equation}
which shows that in incomplete protocols for quantum oblivious transfer, $2p_s+p_r$ can be lower than complete protocols. 
For pure states, we can also obtain a tradeoff relation in terms of the failure probability $p_f$ as
\begin{equation}
2p_s + p_r \ge 2 - F-\sqrt{p_f(1-p_f)}.
\end{equation}



\section{Protocols using symmetric pure states}
\label{sec:sym}

We will now derive the failure and success probability for an honest Bob, and Bob's cheating probability, for protocols using symmetric pure states. 
For $N$ symmetric states $\ket{\psi_j}, ~j=0, 1,\ldots , N-1$, it holds that $\ket\psi_j = U^j\ket{\psi_0}$, where $U$ is a unitary transform which satisfies $U^N=\bf 1$.  

Assume that Alice encodes her two bits in the 
states $\ket{\psi_{00}}, \ket{\psi_{01}}, \ket{\psi_{11}}, \ket{\psi_{10}}$, where $\ket{\psi_{01}}=U\ket{\psi_{00}}, \ket{\psi_{11}} = U^2\ket{\psi_{00}}$ and $\ket{\psi_{10}} = U^3\ket{\psi_{00}}$, where $U^4 = \bf 1$.
Two of the pairwise overlaps between the states are $\braket{\psi_{01}}{\psi_{00}} = f$ and  $\braket{\psi_{11}}{\psi_{00}} = g$. The overlap $f$ is in general complex, while due to the symmetry $g$ has to be real, but can be negative. Since the states are symmetric, this also determines $\braket{\psi_{10}}{\psi_{00}} = f^*$ and all other pairwise overlaps.
An honest Bob chooses between a minimum-error measurement to learn the value of the first bit, and a minimum-error measurement for learning the value of the second bit. The optimal cheating strategy for a dishonest Bob is to make the minimum-error measurement to distinguish between all four equiprobable states.

We will first calculate Bob's cheating probability $p_r$ in this non-interactive quantum protocol. 
Also, if Bob should distinguish between the four states $\ket{\psi_{00}}, \ket{\psi_{01}}, \ket{\psi_{11}}, \ket{\psi_{10}}$ in the last step of a protocol with many steps, then the $p_r$ calculated below will be a lower bound on his cheating probability.
If the states are equiprobable and symmetric, then the optimal measurement a cheating Bob can make is the so-called square-root measurement. Its success probability 
an be obtained in terms of the sum of the square roots of the eigenvalues of the Gram matrix $\mathcal G$ for the states~\cite{Andersson0026}. The Gram matrix has elements $\mathcal{G}_{ij} = \braket{\psi_i}{\psi_j}$.
For the four states we are considering, it is given by 
\begin{equation}
\label{eq:generalgram}
\mathcal{G}=\left(\begin{array}{cccc}1 & f & g & f^* \\f^* & 1 & f & g \\g & f^* & 1 & f \\f & g & f^* & 1\end{array}\right).
\end{equation}
Its eigenvalues are equal to
\begin{eqnarray}
\lambda_0 &=& 1+f+g+f^*,~~ \lambda_1 = 1+if-g-if^*,\nonumber\\
 \lambda_2 &=& 1-f+g-f^*,~~ \lambda_3 = 1-if-g+if^*.
\end{eqnarray}
The number of nonzero eigenvalues of the Gram matrix is equal to the dimension $D$ of the space spanned by the states $\ket{\psi_{00}}, \ket{\psi_{01}}, \ket{\psi_{11}}, \ket{\psi_{10}}$. Independent of $D$,
the success probability for the square-root measurement for symmetric states can be obtained as
\begin{eqnarray}
\label{eq:symbobcheat}
p_r &=& \frac{1}{16}\left|\sqrt{\lambda_0}+\sqrt{\lambda_1}+\sqrt{\lambda_2}+\sqrt{\lambda_3} \right|^2\\
&=&\frac{1}{16}\left|\sqrt{1+g + 2 \Re f}+\sqrt{1+g - 2 \Re f} \right.\nonumber\\
&&+\left.\sqrt{1-g + 2 \Im f}+\sqrt{1-g - 2 \Im f} \right|^2. \nonumber
\end{eqnarray}
This is Bob's optimal cheating probability in a non-interactive protocol using symmetric pure states.

To determine the value of the first bit, an honest Bob needs to distinguish between the equiprobable states
 \begin{equation}
\frac{1}{2}\left(\sigma_{00}+ \sigma_{01}\right){\rm  ~~and~~ } \frac{1}{2}(\sigma_{10}+ \sigma_{11}),
 \end{equation}
 where $\sigma_{ij} = \ket{\psi_{ij}}\bra{\psi_{ij}}$.
  To determine the second bit, he needs to distinguish between the equiprobable states 
\begin{equation}
\frac{1}{2}(\sigma_{00}+\sigma_{10}){~~~\rm and~~~}  \frac{1}{2}(\sigma_{01}+\sigma_{11}).
\end{equation}
Since all states $\ket{\psi_{ij}}$ are equiprobable and symmetric, his failure probability $p_f$ is the same in each case. For determining the first bit, the failure probability is given by
\begin{equation}
p_{f,0} = \frac{1}{2}(1-\frac{1}{4}{\rm Tr}|\sigma_{00}+\sigma_{01}-\sigma_{10}-\sigma_{11}|).
\end{equation}
The analogous expression for the failure probability for determining the second bit, $p_{f,1}$, is obtained by swapping $\sigma_{01}$ and $\sigma_{10}$. The failure probability $p_f=p_{f,0}=p_{f,1}$ for symmetric pure states is calculated in Appendix \ref{app:failp}, and is given by
\begin{widetext}
\begin{eqnarray}
\label{eq:sympfail}
p_f = \frac{1}{2} \bigg[1&-&\frac{1}{4}
\sqrt{(\lambda_0+\lambda_2)(\lambda_1+\lambda_3)
+ \sqrt {(\lambda_0+\lambda_2)^2(\lambda_1+\lambda_3)^2 - 16\lambda_0\lambda_1\lambda_2\lambda_3}}\nonumber\\
&-&\frac{1}{4}
\sqrt{(\lambda_0+\lambda_2)(\lambda_1+\lambda_3)
- \sqrt {(\lambda_0+\lambda_2)^2(\lambda_1+\lambda_3)^2 - 16\lambda_0\lambda_1\lambda_2\lambda_3}}
 \bigg]\nonumber\\
 =\frac{1}{2}\bigg[1&-&
\frac{1}{2} \sqrt{1-g^2+ 2\sqrt{(1+g)^2(\Im f)^2 + (1-g)^2(\Re f)^2 - 4(\Re f)^2(\Im f)^2}}\nonumber\\
&-&
\frac{1}{2} \sqrt{1-g^2- 2\sqrt{(1+g)^2(\Im f)^2 + (1-g)^2(\Re f)^2 - 4(\Re f)^2(\Im f)^2}} \bigg]\nonumber\\
 =\frac{1}{2}\bigg[1&-&
\frac{1}{2} \sqrt{1-g^2+ 2\sqrt{|f|^2(1+g^2) + 2g[(\Im f)^2-(\Re f)^2] - 4(\Re f)^2(\Im f)^2}}\nonumber\\
&-&
\frac{1}{2} \sqrt{1-g^2- 2\sqrt{|f|^2(1+g^2) + 2g[(\Im f)^2-(\Re f)^2] - 4(\Re f)^2(\Im f)^2}}
 \bigg].
\end{eqnarray}
This is the probability that an honest Bob obtains an incorrect bit value.
\end{widetext}

Using the expressions in \eqref{eq:symbobcheat} and \eqref{eq:sympfail}, one can investigate how low $p_r$ can be for a given $p_f$, whether quantum protocols using symmetric pure states can be better than classical protocols, and what the corresponding sets of pure symmetric states are. 
In Appendix \ref{app:symopt}, we show that in the range $(1-1/\sqrt 2)/2 \le p_f \le 1/2$, the lowest possible cheating probability $p_r$ for protocols using symmetric pure states is given by
\begin{eqnarray}
\label{eq:qubitoptBOT}
p_r &=&\frac{1}{4}(1+\sqrt{2})-\frac{1}{\sqrt{2}}p_f,\nonumber\\
 &=& \frac{1}{4}(1-\sqrt{2})+\frac{1}{\sqrt{2}}(1-p_f),
\end{eqnarray}
which we also wrote in terms of the protocol success probability $1-p_f$. Bob's cheating probability is then in the range $1/4 \le p_r \le 1/2$. Moreover, we learn from the analysis in Appendix \ref{app:symopt} that the corresponding optimal states $\ket{\psi_{ij}}$  span a two-dimensional space, since two of the eigenvalues of their Gram matrix are equal to zero. If all four states are equal, then $p_r=1/4$ and $p_f=1/2$. 
The value $p_f = (1-1/\sqrt 2)/2\approx0.15$ corresponds to $p_r=1/2$ (which Bob could also achieve with a random guess).

Let us check when Bob's cheating probability for this class of optimal quantum protocols is lower than in the corresponding classical protocols.
For a quantum protocol to be better than classical protocols, it should hold that $p_r < 1-2p_f$. Combined with \eqref{eq:qubitoptBOT}, this gives $p_f < (3-\sqrt{2})/(8-2\sqrt{2})\approx 0.31$, corresponding to $1-p_f > 0.69$. In this range, Bob's cheating probability is therefore lower in the quantum protocols. When $p_f \gtrsim 0.31$, protocols using symmetric pure states are not optimal, which is also interesting. Mixedness and/or asymmetry must then be useful, and might of course be useful also for other values of $p_f$. In the classical protocols, Alice uses the equivalent of mixed states. 

\begin{figure}
\includegraphics[scale=1]{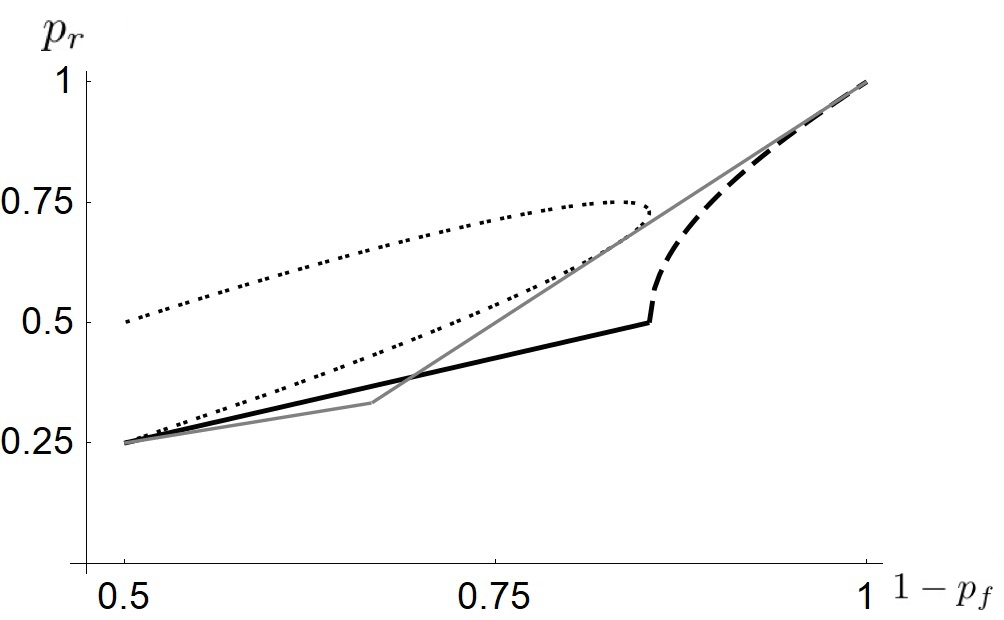} 
\caption{The lowest possible cheating probability $p_r$ for the receiver Bob, as a function of the protocol success probability $1-p_f$, in non-interactive protocols for oblivious transfer. The sender Alice can only cheat with probability $p_s=1/2$ (a random guess). The grey line shows the lowest possible $p_r$ in classical protocols. The solid black line and dashed black line show the lowest possible $p_r$ for quantum protocols using symmetric pure states, in the ranges $1/4\le p_r\le 1/2$ and $1/2 \le p_r \le 1$, respectively. For the solid black line, the states $\ket{\psi_{ij}}$ span a two-dimensional space, and for the dashed line, a four-dimensional space. The dotted line shows the cheating probability achieved by the family of states in section \ref{sec:qutrits}, spanning a three-dimensional space.}
\label{fig:optsym}
\end{figure}

When $0\le p_f \le (1-1/\sqrt 2)/2\approx0.15$, which corresponds to $1/2\le p_r \le 1$, the lowest possible $p_f$ is given by
\begin{equation}
\label{eq:ququartoptPf}
p_f = \frac{1}{2}\left[1-\sqrt{p_r^2+(1-p_r)^2} \right].
\end{equation}
The optimisation in Appendix \ref{app:symopt} also tells us that the states $\ket{\psi_{ij}}$ now span a four-dimensional space.
Geometrically, if $p_r$ and $1-p_r$ are the lengths of two sides of a right-angled triangle, then the length of the hypotenuse is given by $1-2p_f$. Hence, $p_r$ is always less than $1-2p_f$, meaning that for these optimal quantum protocols using symmetric pure states, Bob always has a lower cheating probability than in classical protocols with the same failure probability $p_f$. To summarize, protocols using symmetric pure quantum states are better than corresponding classical protocols when $0<p_f \lesssim 0.31$. When $0.31\lesssim p_f < 1/2$, classical protocols are better, meaning that mixedness and/or asymmetricity must be useful in more general quantum protocols (which must be at least as good as classical protocols). The optimal $p_r$ for symmetric pure states is plotted in figure \ref{fig:optsym} as a function of the protocol success probability $1-p_f$.

\section{Optimal protocols using symmetric pure states}
\label{sec:opt}

In this section we will describe sets of four symmetric pure states which are optimal in the sense that for a given failure probability $p_f$, Bob's cheating probability $p_r$ is as low as possible. Alternatively, for a given $p_r$, Bob's $p_f$ is as low as possible. 
We also describe a set of qutrit states which is suboptimal, but is related to the states used in the semi-random oblivious transfer protocol in~\cite{Ryan}.

\subsection{BB84 or ``Wiesner" states}

Suppose that Alice encodes her two bits $x_0, x_1$  in the states
\begin{eqnarray}
\ket{\psi_{00}}&=&\ket 0, ~~~\ket{\psi_{01}}=\sqrt{i}\ket +,\\ 
\ket{\psi_{11}}&=&i\ket 1, ~~\ket{\psi_{10}}=i\sqrt{i}\ket -,
\end{eqnarray}
where $\ket{+}= (\ket{0}+\ket{1})/\sqrt{2}$ and $\ket{-}= (\ket{0}-\ket{1})/\sqrt{2}$. Apart from the overall phase factors which have been included to make the states symmetric according to the definition at the start of Section \ref{sec:sym}, this set of states is identical to the one used for ``conjugate coding" by Wiesner~\cite{Wie83}, in the Bennett-Brassard protocol for quantum cryptography (BB84)~\cite{BB84}, and for quantum random access codes~\cite{QRAC}. Here we only reinterpret the situation as imperfect 1-out-of-2 oblivious transfer. 
Depending on which bit value an honest Bob wishes to obtain, he measures in a basis rotated by $\pi/8$ from the basis $\{\ket{0},\ket{1}\}$ to obtain the first bit, or rotated by $3\pi/8$ to obtain the second bit.
Either of these measurements succeed with probability $\cos^2{(\frac{\pi}{8})} \approx 0.85$. Consequently, the protocol's inherent failure probability $p_f^W$ is given by
\begin{equation}
p_f^W = 1-\cos^2{(\frac{\pi}{8})}\approx 0.15.
\end{equation}
As before, no-signalling prohibits Alice from cheating so that $p_s=1/2$. The success probability of Bob's minimum-error measurement for distinguishing between all four states is
\begin{equation}
p_r^W = 1/2,
\end{equation}
which is his cheating probability.
It holds that $p_r^W < 1-2p_f^W \approx 0.7$. This protocol therefore is better than classical protocols. In fact, as is confirmed by the results in the previous section and in Appendix \ref{app:symopt}, it is optimal in the sense that for this value of $p_f$, no choice of four symmetric pure states can give a lower $p_r$.

\subsection{Qubit states}
\label{sec:qubits}

To obtain protocols with $p_f$ in the range $ p_f^W \le p_f \le 1/2$,
we will look at symmetric qubit states given by
\begin{eqnarray}
\label{eq:qubitoptstates} 
\ket{\psi_{00}}&=&a\ket{0}+b\ket{1},
~~\ket{\psi_{01}}=a\ket{0}+ib\ket{1},\nonumber\\
\ket{\psi_{11}}&=&a\ket{0}-b\ket{1},
~~\ket{\psi_{10}}=a\ket{0}-ib\ket{1},
\end{eqnarray}
where 
$0\le |a|, |b| \le 1$, with $|a|^2+|b|^2=1$. This set of states is symmetric with the symmetry operation $U=\ket{0}\bra{0} + i\ket{1}\bra{1}$. For $a=b=1\sqrt{2}$ we obtain the eigenstates of $\sigma_x$ and $\sigma_y$, that is, four states isomorphic to the ``Wiesner" or BB84 states, which are eigenstates of $\sigma_x$ and $\sigma_z$. 
By varying $a$ and $b$, we will generally obtain lower $p_r$ and higher $p_f$ than for the ``Wiesner" states. For $|a|=1, b=0$ or $a=0, |b|=1$, all four states are equal to the same state, giving $p_f=1/2$ and $p_r=1/4$. 

It holds that $f=|a|^2+i|b|^2$ and $g=|a|^2-|b|^2$. From \eqref{eq:symbobcheat} and \eqref{eq:sympfail} we obtain
\begin{eqnarray}
p_f &=& \frac{1}{2}(1-\sqrt{2}|ab|),\nonumber\\
p_r &=& \frac{1}{4}(|a|+|b|)^2 = \frac{1}{4}(1+2|ab|),
\end{eqnarray}
and it therefore also holds that
\begin{equation}
p_r = \frac{1}{\sqrt{2}}(1-p_f)+\frac{1}{4}(1-\sqrt{2}),
\end{equation}
that is, Bob's cheating probability $p_r$ is a straight line as a function of the protocol success probability $1-p_f$. 
As shown in Appendix \ref{app:symopt} and given in equation \eqref{eq:qubitoptBOT},  this is the lowest possible $p_r$ for a given $p_f$ with $p_f$ in the range $p_f^W \le p_f \le 1/2 $. It follows that the qubit states in \eqref{eq:qubitoptstates} are indeed optimal.
Moreover, it holds that 
\begin{equation}
1-2p_f = \sqrt 2|ab| <p_r
\end{equation}
when $|ab|>1/(4\sqrt 2-2) \approx 0.273$, which corresponds to $p_f\lesssim 0.31$. As mentioned above, this is the range where this class of protocols outperform classical protocols.

\subsection{Ququart states}
\label{sec:qutrits}

We will now look at a set of states which makes it possible to reach $p_f=0$ (when it must hold that $p_r=1$). For this limit it is necessary that all four states become orthogonal, and we also know from the optimisation in Appendix \ref{app:symopt} that the corresponding sets of states need to span a four-dimensional space. We will choose
\begin{eqnarray}
\ket{\psi_{00}}&=&\frac{1}{\sqrt 2}(\ket{0}+\ket{1}),
~~\ket{\psi_{01}}=\frac{a}{\sqrt{2}}(\ket{0}+i\ket{1})+b\ket{2},\nonumber\\
\ket{\psi_{11}}&=&\frac{1}{\sqrt 2}(\ket{0}-\ket{1}),
~~\ket{\psi_{10}}=\frac{a}{\sqrt{2}}(\ket{0}-i\ket{1})+b\ket{3},\nonumber\\
~~\end{eqnarray}
where $0\le |a|, |b| \le 1$ and $|a|^2+|b|^2=1$. For $a=1$ we obtain the eigenstates of $\sigma_x$ and $\sigma_y$, connecting with the qubit 
states considered above. The pairwise overlaps are now 
\begin{eqnarray}
\braket{\psi_{00}}{\psi_{01}}= \braket{\psi_{11}}{\psi_{10}} &=&\frac{a}{2}(1+i),  \nonumber\\
~~\braket{\psi_{01}}{\psi_{11}}= \braket{\psi_{10}}{\psi_{00}}&=&\frac{a^*}{2}(1+i) ,\nonumber\\
\braket{\psi_{00}}{\psi_{11}}= \braket{\psi_{01}}{\psi_{10}} &=&0.
\end{eqnarray}
That is, the set of sates is symmetric without further modifications if $a$ is real. In this case we obtain $f=(a/2)(1+i)$ and $g=0$, and 
\begin{eqnarray}
p_f &=& \frac{1}{4}\left(2-\sqrt{1+a\sqrt{2-a^2}}-\sqrt{1-a\sqrt{2-a^2}} \right),
\nonumber\\
p_r &=& \frac{1}{4}(\sqrt{1+a}+\sqrt{1-a})^2 = \frac{1}{2}(1+\sqrt{1-a^2}).
\end{eqnarray}
The above relations imply that
\begin{equation}
(1-2p_f)^2=p_r^2+(1-p_r)^2=1-\frac{a^2}{2}.
\end{equation} 
This agrees with equation \eqref{eq:ququartoptPf}, and the above choice of ququart states is therefore optimal for $0\le p_f \le p_f^W$. For this set of states, it holds that $1-2p_f \ge  p_r$ for the whole range of $0\le a \le 1$, with equality for $a=0$, when $p_f=0$ and $p_r=1$. The resulting quantum protocol therefore outperforms classical protocols for the whole range of $0 < p_f \le p_f^W$.

\subsection{Qutrit states}

As an example of symmetric pure states which turn out to not be optimal for the type of protocols we are considering, we will look at the set of states
\begin{eqnarray}
\label{eq:qutritstates}
\ket{\psi_{00}}&=&a\ket{0}+b\ket{1},
~~\ket{\psi_{01}}=a\ket{0}+b\ket{2},\nonumber\\
\ket{\psi_{11}}&=&a\ket{0}-b\ket{1},
~~\ket{\psi_{10}}=a\ket{0}-b\ket{2},
\end{eqnarray}
where $0\le |a|, |b| \le 1$ and $|a|^2+|b|^2=1$. 
One reason to investigate this set of states is that
for $a=b=1/\sqrt{2}$, it is 
equivalent to two copies of the same ``Wiesner" states, that is, the states $\ket{00}, \ket{++}, \ket{11}, \ket{- -}$, since all pairwise overlaps match. These states are used in an oblivious transfer protocol in~\cite{Ryan}, which has the lowest known cheating probabilities for a complete 
1-out-of-2 quantum oblivious transfer protocol where the parties are not restricted. It is therefore interesting to see how well the same set of states performs for an incomplete oblivious transfer protocol.

If Bob is given two copies of a ``Wiesner" or BB84 state, we might expect that an honest Bob will obtain a correct bit value more often than if he is given a single copy, but also, that it becomes easier for a dishonest Bob to cheat. However, one can easily verify that this is not the case. While an additional copy helps a dishonest Bob cheat, as we will see, {\em it does not lower an honest Bob's failure probability}. An honest Bob can ignore the second copy. On the other hand, with two copies of a ``Wiesner" or BB84 state, the semi-random oblivious transfer protocol with $p_f=0$ in \cite{Ryan} is possible. In a semi-random protocol, Bob does not choose whether he obtains the first or the second bit value, but obtains a bit value at random. Bob's overall measurement is then no longer from Alice's point of view equivalent to Bob choosing between two measurements. Instead, Bob makes one fixed measurement with four outcomes. Therefore while $p_f=0$ becomes possible, Bob's and also Alice's cheating probability increases, becoming equal to $\approx 0.729$ and  $3/4$, respectively, if the states $\ket{00}, \ket{++}, \ket{11}, \ket{- -}$ are used.

Returning to non-random protocols where Bob chooses between learning the first or the second bit, 
an honest Bob does not benefit from using the qutrit states in equation \eqref{eq:qutritstates}, as compared with the qubit states in equation \eqref{eq:qubitoptstates}. This may seem counter-intuitive, but is analogous to a scenario where Bob is sent a classical bit value that might have been flipped with some probability $p$. His probability to read out the bit value correctly does not increase if he is sent a second copy of the same bit value, which might also independently have been flipped with probability $p$. Both bits are then incorrect (and the two incorrect values agree) with probability $p^2$. If one bit value is correct and the other one not, then Bob will have to guess what the correct value is. His overall probability to arrive at the incorrect bit value is then $p^2+2p(1-p)/2=p$, the same as if he had been sent only one bit.

For the states in \eqref{eq:qutritstates}, we obtain $f=\braket{\psi_{00}}{\psi_{01}}=|a|^2$ and $g=\braket{\psi_{00}}{\psi_{11}} = |a|^2-|b|^2$. This gives
\begin{eqnarray}
p_f &=& \frac{1}{2}(1-\sqrt{2}|ab| )\\
p_r&=&\frac{1}{4}(|a|+\sqrt{2}|b|)^2 = \frac{1}{4}(1+|b|^2+2\sqrt{2}|ab|).\nonumber
\end{eqnarray}
That is, as a function of $|a|$ and $|b|$, we obtain the same $p_f$ as for the qubit states in \eqref{eq:qubitoptstates}, but a higher $p_r$. This set of qutrit states therefore never outperforms the qubit states. Bob's cheating probability for this set of qutrit states is plotted in figure \ref{fig:optsym}, along with the lowest possible cheating probabilities for the optimal qubit and ququart states, and the lowest possible cheating probability for classical protocols.


\section{Conclusion and Outlook}

We have demonstrated that for oblivious transfer protocols which sometimes fail, quantum protocols may outperform classical ones, in the sense that for protocols where the sender can only cheat with probability $1/2$, Bob's cheating probability in quantum protocols can be lower than in any classical non-interactive protocol with the same failure probability $p_f$.
We fully characterized non-interactive quantum protocols which use symmetric pure states, and gave examples of optimal  protools, which use qubits and ququarts respectively. When $p_f \lesssim 0.31$, Bob's cheating probability in these protocols is lower than in classical protocols.

Incomplete 1-out-of-2 oblivious transfer of the type we are considering, where the sender can only cheat with probability $p_s=1/2$, can be seen as a generalisation of complete oblivious transfer, but also as a generalization of random access codes (RACs)~\cite{Wie83, QRAC}. For a RAC, in the simplest case (which will correspond most closely to 1-out-of-2 oblivious transfer), one restricts the sender to use one bit to encode the values of two bits. The sender does not know which bit the receiver wants, but aims for the receiver to correctly retrieve the bit of their choice with a probability that is as high as possible. For a corresponding quantum random access code (QRAC)~\cite{Wie83, QRAC}, the sender is restricted to using one qubit for encoding two classical bit values. 
Probabilistic one-time programs~\cite{OTP, Fitz} are a related functionality, which reduces to a QRAC in the case relevant here.

Unlike in oblivious transfer, in a RAC or QRAC one is not concerned with the probability that the sender guesses which bit value the receiver wants to access. But since the receiver is choosing between two measurements, depending on which bit value they choose, the sender is not able to tell better than with a random guess. In the type of oblivious transfer protocols we have been considering, however, the sender Alice can correctly guess which bit the receiver Bob wants only with probability 1/2, which is analogous to a RAC or QRAC. 

Another difference between RACS and QRACs on one hand, and oblivious transfer on the other hand, is that for oblivious transfer, 
there is no restriction on the dimensionality of the state space that can be used. Instead of fixing the dimensionality of the state space, one wants to restrict the probability for the receiver to retrieve all of the sent information.  In this sense, the type of incomplete oblivious transfer we have considered, with $p_s =1/2$, is a generalisation of a QRAC (as well as being a generalization of complete or ``perfect" oblivious transfer).
That is, for the type oblivious transfer we have considered, we are concerned with (i) maximising the probability that the receiver correctly obtains their chosen bit value, as in a RAC, but also (ii) minimising the probability that they correctly obtain both bit values, which generalizes the restriction of sending a single bit or qubit. 
A remaining difference is that in complete protocols for oblivious transfer, the receiver should always correctly retrieve their chosen bit value, whereas in a RAC or QRAC, or in an incomplete protocol for oblivious transfer, the probability for the bit value to be correct is generally lower than 1.  Our protocols interpolate between or ``connect" random access codes and oblivious transfer, generalizing both functionalities.

To summarize, we have obtained bounds on cheating probabilities for incomplete (imperfect) quantum oblivious transfer, and derived optimal protocols which use pure symmetric states. Cheating probabilities can be lower for incomplete protocols than for complete protocols. The optimal symmetric-state protocols use qubit or ququart (2-qubit) states, and could be implemented with existing experimental techniques; ququart states can for example be realized as single-photon states, using either four different spatial modes, or polarisation and two spatial modes. The honest receiver's measurements are easy-to-realize projective measurements, since the task is to distinguish between only two mixed states. As is well known and also discussed above, the optimal measurement is in this case simply a projective measurement in a particular basis. The type of incomplete oblivious transfer we have considered can be seen as generalizing both complete oblivious transfer and random access codes, and connecting these two functionalities.

\acknowledgments

IVP and EA acknowledge support by the UK Engineering and Physical Sciences Research Council (EPSRC) under Grants No. EP/T001011/1.


\appendix
\begin{widetext}
\section{Protocol failure probability for symmetric pure states}
\label{app:failp}

To calculate the protocol failure probability, we construct the states
\begin{eqnarray}
2\sqrt{\lambda_0}\ket{A_0} &=& \ket{\psi_{00}}+\ket{\psi_{01}}+\ket{\psi_{11}}+\ket{\psi_{10}}\nonumber\\
2\sqrt{\lambda_1}\ket{A_1} &=& \ket{\psi_{00}}+i\ket{\psi_{01}}-\ket{\psi_{11}}-i\ket{\psi_{10}}\nonumber\\
2\sqrt{\lambda_2}\ket{A_2} &=& \ket{\psi_{00}}-\ket{\psi_{01}}+\ket{\psi_{11}}-\ket{\psi_{10}}\nonumber\\
2\sqrt{\lambda_3}\ket{A_3} &=& \ket{\psi_{00}}-i\ket{\psi_{01}}-\ket{\psi_{11}}+i\ket{\psi_{10}},
\end{eqnarray}
where $\lambda_i$ are the previously given eigenvalues of the Gram matrix for the states $\ket{\psi_{ij}}$. If the space spanned by $\ket{\psi_{00}}, \ket{\psi_{01}}, \ket{\psi_{11}}, \ket{\psi_{10}}$ is four-dimensional, then the states $\ket{A_0}, \ket{A_1}, \ket{A_2}, \ket{A_3}$ are orthonormal. If the space is less than four-dimensional, then some of the eigenvalues $\lambda_i$ are equal to zero, but the four states $\ket{A_i}$ can still be chosen orthonormal, and used as a basis. Independent of dimension, we have
\begin{eqnarray}
\ket{\psi_{00}} &=& \frac{1}{2}\left(\sqrt{\lambda_0}\ket{A_0} +\sqrt{\lambda_1}\ket{A_1} +\sqrt{\lambda_2}\ket{A_2} +\sqrt{\lambda_3}\ket{A_3} \right)\nonumber\\
\ket{\psi_{01}} &=& \frac{1}{2}\left(\sqrt{\lambda_0}\ket{A_0} -i\sqrt{\lambda_1}\ket{A_1} -\sqrt{\lambda_2}\ket{A_2} +i\sqrt{\lambda_3}\ket{A_3} \right)\nonumber\\
\ket{\psi_{11}} &=& \frac{1}{2}\left(\sqrt{\lambda_0}\ket{A_0} -\sqrt{\lambda_1}\ket{A_1} +\sqrt{\lambda_2}\ket{A_2} -\sqrt{\lambda_3}\ket{A_3} \right)\nonumber\\
\ket{\psi_{10}} &=& \frac{1}{2}\left(\sqrt{\lambda_0}\ket{A_0} +i\sqrt{\lambda_1}\ket{A_1} -\sqrt{\lambda_2}\ket{A_2} -i\sqrt{\lambda_3}\ket{A_3} \right).\nonumber\\
\end{eqnarray}
Using the basis $\{\ket{A_0}, \ket{A_1}, \ket{A_2}, \ket{A_3}\}$ we therefore have
\begin{eqnarray}
&&\frac{1}{2}\left(\sigma_{00}+\sigma_{01}\right)=
\frac{1}{8}\left(\begin{array}{cccc}
2\lambda_0 & \sqrt{\lambda_{01}}(1+i) & 0 & \sqrt{\lambda_{03}}(1-i) \\
\sqrt{\lambda_{01}}(1-i) & 2\lambda_1 & \sqrt{\lambda_{12}}(1+i) & 0 \\
0 & \sqrt{\lambda_{12}}(1-i) & 2\lambda_2 & \sqrt{\lambda_{23}}(1+i) \\
\sqrt{\lambda_{03}}(1+i) & 0 & \sqrt{\lambda_{23}}(1-i) & 2\lambda_3
\end{array}\right),\nonumber\\
\\
&&\frac{1}{2}\left(\sigma_{10}+\sigma_{11}\right)=
\frac{1}{8}\left(\begin{array}{cccc}
2\lambda_0 & \sqrt{\lambda_{01}}(-1-i) & 0 & \sqrt{\lambda_{03}}(-1+i) \\
\sqrt{\lambda_{01}}(-1+i) & 2\lambda_1 & \sqrt{\lambda_{12}}(-1-i) & 0 \\
0 & \sqrt{\lambda_{12}}(-1+i) & 2\lambda_2 & \sqrt{\lambda_{23}}(-1-i) \\
\sqrt{\lambda_{03}}(-1-i) & 0 & \sqrt{\lambda_{23}}(-1+i) & 2\lambda_3
\end{array}\right),\nonumber\\
\\
{\rm and}~~~&&\frac{1}{4}\left(\sigma_{00}+\sigma_{01}-\sigma_{10}-\sigma_{11}\right)=
\frac{1}{8}\left(\begin{array}{cccc}
0 & \sqrt{\lambda_{01}}(1+i) & 0 & \sqrt{\lambda_{03}}(1-i) \\
\sqrt{\lambda_{01}}(1-i) & 0& \sqrt{\lambda_{12}}(1+i) & 0 \\
0 & \sqrt{\lambda_{12}}(1-i) & 0 & \sqrt{\lambda_{23}}(1+i) \\
\sqrt{\lambda_{03}}(1+i) & 0 & \sqrt{\lambda_{23}}(1-i) & 0
\end{array}\right),\nonumber\\
\label{eq:helmatrix}
\end{eqnarray}
where we have used the shorthand $\lambda_{ij} = \lambda_i\lambda_j$. 
The eigenvalue equation for the last matrix above is given by $|\frac{1}{4}\left(\sigma_{00}+\sigma_{01}-\sigma_{10}-\sigma_{11}\right) - \Lambda {\bf 1}|=0$, which, when evaluating the determinant, gives
\begin{equation}
(8\Lambda)^4-2(8\Lambda)^2(\lambda_0+\lambda_2)(\lambda_1+\lambda_3)+16\lambda_{0123}=0,
\end{equation}
where $\lambda_{0123} = \lambda_0\lambda_2\lambda_2\lambda_3$.
The four eigenvalues of the matrix in \eqref{eq:helmatrix} are consequently given by the solutions to
\begin{eqnarray}
\Lambda^2 &=& \frac{1}{8^2}\big[( \lambda_0+\lambda_2)(\lambda_1+\lambda_3)
\pm \sqrt {(\lambda_0+\lambda_2)^2(\lambda_1+\lambda_3)^2 - 16\lambda_0\lambda_1\lambda_2\lambda_3}\big]\nonumber\\
&=&\frac{1}{8^2}\big[4(1-g^2)\pm 8\sqrt{(1+g)^2(\Im f)^2 + (1-g)^2(\Re f)^2 - 4(\Re f)^2(\Im f)^2}\big].\nonumber
\end{eqnarray}
The failure probability for the protocol is therefore given by
\begin{eqnarray}
\label{eq:appsympfail}
p_f = \frac{1}{2} \bigg[1&-&\frac{1}{4}
\sqrt{(\lambda_0+\lambda_2)(\lambda_1+\lambda_3)
+ \sqrt {(\lambda_0+\lambda_2)^2(\lambda_1+\lambda_3)^2 - 16\lambda_0\lambda_1\lambda_2\lambda_3}}\nonumber\\
&-&\frac{1}{4}
\sqrt{(\lambda_0+\lambda_2)(\lambda_1+\lambda_3)
- \sqrt {(\lambda_0+\lambda_2)^2(\lambda_1+\lambda_3)^2 - 16\lambda_0\lambda_1\lambda_2\lambda_3}}
 \bigg]\nonumber\\
 =\frac{1}{2}\bigg[1&-&
\frac{1}{2} \sqrt{1-g^2+ 2\sqrt{|f|^2(1+g^2) + 2G[(\Im f)^2-(\Re f)^2] - 4(\Re f)^2(\Im f)^2}}\nonumber\\
&-&
\frac{1}{2} \sqrt{1-g^2- 2\sqrt{|f|^2(1+g^2) + 2G[(\Im f)^2-(\Re f)^2] - 4(\Re f)^2(\Im f)^2}}
 \bigg].
\end{eqnarray}
This is the probability that an honest Bob obtains an incorrect bit value.

\end{widetext}

\section{Optimal pure-state protocols}
\label{app:symopt}

For an optimal protocol, it holds that for a given failure probability $p_f$, Bob's cheating probability $p_r$ is as low as possible, and vice versa. The failure probability $p_f$ and the cheating probability $p_r$ are given in equations \eqref{eq:sympfail} (which is the same as \eqref{eq:appsympfail})  and \eqref{eq:symbobcheat} respectively. We will minimize $p_f$ for a given (fixed) $p_r=|\sum_i\sqrt{\lambda_i}|^2/16$. We first rewrite equation  \eqref{eq:appsympfail} in the more convenient form
\begin{equation}
\label{eq:FPf}
8(1-2p_f)^2 = (\lambda_0+\lambda_2)(\lambda_1+\lambda_3)+4\sqrt{\lambda_{0123}}.
\end{equation}
In the range $0\le p_f\le 1/2$, minimising $p_f$ is equivalent to maximising $(1-2p_f)^2$. It holds that
\begin{equation}
\lambda_0+\lambda_1+\lambda_2+\lambda_3=4,
\end{equation}
where we note that $\lambda_i$ are all real, and $\lambda_i\ge 0$.
We define
\begin{eqnarray}
a=(\sqrt{\lambda_0}+\sqrt{\lambda_1}+\sqrt{\lambda_2}+\sqrt{\lambda_3})/4\nonumber\\
b=(\sqrt{\lambda_0}-\sqrt{\lambda_1}+\sqrt{\lambda_2}-\sqrt{\lambda_3})/4\nonumber\\
c=(\sqrt{\lambda_0}+\sqrt{\lambda_1}-\sqrt{\lambda_2}-\sqrt{\lambda_3})/4\nonumber\\
d=(\sqrt{\lambda_0}-\sqrt{\lambda_1}-\sqrt{\lambda_2}+\sqrt{\lambda_3})/4,
\end{eqnarray}
which means that $a^2=p_r$ is a constant. We also obtain
\begin{equation}
a^2+b^2+c^2+d^2=\frac{1}{4}(\lambda_0+\lambda_1+\lambda_2+\lambda_3)=1,
\end{equation}
which together means that
\begin{equation}
b^2+c^2+d^2=1-p_r
\end{equation}
is also a constant. Equation \eqref{eq:FPf} becomes
\begin{equation}
(1-2p_f)^2=(a^2-b^2)^2+(c^2-d^2)^2,
\end{equation}
which should be maximized. Recall that $a^2=p_r$ and  $b^2+c^2+d^2=1-p_r$ are constant. Clearly, $(1-2p_f)^2$ will be maximized if we can choose $b=0$, and either $c=0$ and $d^2=1-p_r$, or $c^2=1-p_r$ and $d=0$. If this is possible, then we obtain
\begin{equation}
(1-2p_f)^2=p_r^2+(1-p_r)^2.
\end{equation}
It is however not always possible to choose $b=0$, and either $c=0$ or $d=0$. For example, if $b=d=0$, then $a=\sqrt{p_r}$ and $c=\sqrt{1-p_r}$, and we obtain
\begin{eqnarray}
\sqrt{\lambda_0}=a+b+c+d=\sqrt{p_r}+\sqrt{1-p_r}\nonumber\\
\sqrt{\lambda_1}=a-b+c-d=\sqrt{p_r}+\sqrt{1-p_r}\nonumber\\
\sqrt{\lambda_2}=a+b-c-d=\sqrt{p_r}-\sqrt{1-p_r}\nonumber\\
\sqrt{\lambda_3}=a-b-c+d=\sqrt{p_r}-\sqrt{1-p_r}.
\end{eqnarray}
Since $\lambda_i\ge 0$, it must hold that $1/2\le p_r\le 1$ for this to be possible. This corresponds to $0\le p_f\le (1-1/\sqrt{2})/2\approx 0.15$, where it then holds that
\begin{equation}
p_f = \frac{1}{2}\left[1-\sqrt{p_r^2+(1-p_r)^2} \right].
\end{equation}
When $1/2 < p_r$, the space spanned by the corresponding optimal states $\ket{\psi_{ij}}$ is four-dimensional, and becomes two-dimensional for $p_r=1/2$ (equivalent to the ``Wiesner" states).

When $1/4\le p_r < 1/2$, we must instead choose $a^2=c^2=p_r$ and $b^2=d^2=1/2-p_r$ to maximize $(1-2p_f)^2$ (equivalently, to minimize $p_f$). Alternatively, we can choose $a^2=d^2$ and $b^2=c^2$. It will either hold that  $\lambda_1=\lambda_2=0$ or that $\lambda_2=\lambda_3=0$, meaning that the corresponding optimal sets of states $\ket{\psi_{ij}}$ span a two-dimensional space. It then holds that
\begin{equation}
p_r=\frac{1}{4}(1+\sqrt{2})-\frac{1}{\sqrt{2}}p_f,
\end{equation}
which is the lowest possible cheating probability $p_r$ in the range $(1-1/\sqrt{2})/2 \le p_f\le 1/2$, corresponding to  $1/4\le p_r\le 1/2$, for protocols using symmetric pure states. The results in this Appendix have also been verified numerically.

\end{document}